\documentclass[final]{aipproc}
\layoutstyle{6x9}
\usepackage{amsmath,bm,graphicx}

 \newcommand{\Half}{\textstyle\frac{1}{2}}
 \newcommand{\D}{\textstyle{\rm d}}
 \newcommand{\E}{\textstyle{\rm e}}
 \newcommand{\I}{\textstyle{\rm i}}

\begin{document}
\title{Two-proton correlation function: a gentle introduction}


\author{A. Deloff}{address={
Institute for Nuclear Studies, Warsaw}}

\begin{abstract}
The recent COSY-11 collaboration measurement
of the two-proton correlation function
in the $pp\to pp\eta$ reaction, reported
 at this meeting \cite{exp}, arouse some interest in a 
simple theoretical description of the correlation function.
In these notes we present a pedagogical introduction 
to the practical methods that can be used for 
 calculating the correlation function. 
\end{abstract}

\maketitle

 We are going here to deal with
 low-energy phenomena and in order to avoid
unnecessary complications our approach will be  non-relativistic. 
We wish to develop a practical scheme for calculating the two-particle 
correlation function $C(k)$, with $k$ denoting their relative momentum, 
choosing as the departure point the familiar formula 
\begin{equation}
 C(k) = \int D(r)\mid\negthickspace \Phi_{\bm{k}}(\bm{r})
                      \negthickspace\mid^2  \D^3 r,
\label{s1}
\end{equation}
(referred to in the literature as the Koonin-Pratt model \cite{KP}),
expressing $C(k)$ as an overlap of two distributions.
The first distribution, $D(r)$, is the {\it effective} source 
density function
that is usually for convenience assumed to be a Gaussian
\begin{equation}
 D(r)= (4\pi d^2)^{-3/2}\,\E^{-(r/2d)^2} 
 \label{s2}
\end{equation}
with a single parameter $d$ reflecting the size of the source. 
The second factor  
 entering the overlap integral (\ref{s1}),
is a probability density involving the square of 
the wave function $ \Phi_{\bm{k}}(\bm{r}) $ describing
 two-particle system in the continuum.  
\par
Actually, formula (\ref{s1}) is only a static approximation to the correlation
function derived under the assumption that the final-state interaction
between the two detected particles dominates, while all other
interactions are negligible. Furthermore, it is assumed that
the correlation functions are determined by the two-body densities of states
(corrected for their mutual interactions) and that the single particle
phase space distribution function of the emitted particle varies slowly
with momentum. Admittedly, the question concerning 
the validity of these assumptions is
far from settled but since they result in  a manageable calculational
scheme it is worthwhile to examine in some detail its consequences.  
\par 
 Adopting hereafter units in which
 $\hbar=c=1$, for non-interacting particles
 $\Phi_{\bm{k}}(\bm{r})$  takes the form of a  plane wave
 \begin{equation*}
\Phi_{\bm{k}}(\bm{r}) = \E^{\I\bm{k}\cdot\bm{r}}
\end{equation*}
and in this case the correlation function is equal to unity,
which means no correlation. However, when
the particles are non-interacting identical bosons(fermions), the plane wave 
needs to be properly symmetrized(antisymmetrized), viz. 
\begin{equation*}
 \Phi_{\bm{q}}^\pm(\bm{r}) = \dfrac{\E^{\I\bm{k}\cdot\bm{r}} \pm
 \E^{-\I\bm{k}\cdot\bm{r}}}
   {\sqrt{2}}
\end{equation*}
and the interference term yields a non-vanishing contribution
to the correlation function in excess of unity. For a Gaussian source, direct
integration gives
\begin{equation}
C_\pm(k)= 1 \pm \E^{-4d^2k^2}
 \label{s2a}
\end{equation} 
 and in the symmetric case 
$C_+(k)$ peaks at threshold with $C_+(0)=2$ going eventually to unity at large $k$.
By contrast, in the antisymmetric case $C_-(k)$ vanishes at threshold reaching unity from below for large $k$. The rate at which the correlation function
approaches unity is controlled by $d$, the only parameter in the game. 
Actually, $C_\pm(k)$ could be viewed as a function depending on two
parameters $k$ and $d$ but for free propagation, owing to dimensional scaling,
 the correlation function can only depend upon their product $dk$
bearing a universal character.
For interacting particles, as soon as additional parameters become available,
further dimensionless quantities can be formed and 
$dk$ is no longer the only possible combination.
\par
When the two-particle system has additional degrees of freedom
the even and odd components of the wave function may both enter the correlation
function. This happens in the two-proton case where 
the isospin part of the wave function 
is necessarily symmetric and therefore Pauli principle admits 
in the spin-singlet states only even $\ell$
whereas in spin-triplet states, respectively, odd $\ell$.
In the simplest case of two non-interacting
protons the resulting spin weights are purely statistical,
and we get
\begin{equation}
 C_{pp}(k) = \frac{1}{4}C_+(k)+\frac{3}{4}C_-(k)=1-\frac{1}{2}\E^{-4d^2k^2}
 \label{ss2}
\end{equation}
with the intercept $C_{pp}(0)=\Half$.
\par
Apart from correlations associated with permutation symmetry there would
be also dynamical correlations induced by the two-particle interaction.
To get some insight into the nature of dynamical correlations let us
consider the case of two different particles whose propagation is not free
but distorted by their mutual interaction. 
For simplicity we assume that
the interaction may be represented by a spherically symmetric potential
$V(r)$. The wave function may be expanded in partial waves
\begin{equation}
\Phi_{\bm{k}}(\bm{r})=\sum_{\ell=0}^\infty (2\ell+1)\I^\ell\, 
 u_\ell(k,r)/(kr)\,
P_\ell(\hat{\bm{k}}\cdot\hat{\bm{r}})
 \label{s2b}
\end{equation}  
where $P_\ell(z)$ denotes Legendre polynomial and the function $u_\ell(k,r)$
is the solution of the appropriate radial Schr\"{o}dinger equation and
 for $r\to\infty$   satisfies 
the outgoing wave boundary condition
\begin{equation}
 u_\ell(k,r)\sim \sin(kr-\Half\ell\pi) + kf_\ell(k)\,\exp{[\I (kr-\Half\ell\pi)]}
  \label{s2d}
\end{equation} 
where  $f_\ell(k)$ is
the partial wave scattering amplitude. 
 Inserting (\ref{s2b}) in  (\ref{s1}),  after trivial angular integration,
the correlation function takes the form of a series 
\begin{equation}
C(k)=\dfrac{4\pi}{k^2}\sum_{\ell=0}^\infty (2\ell+1)\,\int_0^\infty D(r)
 \mid\negthickspace u_\ell(k,r)\negthickspace\mid^2 \D r
 \label{s2c}
\end{equation}
whose convergence rate depends upon the case considered. For a short ranged potential
the functions $\mid\negthickspace  u_\ell(k,r) \negthickspace\mid^2 $ 
 are pushed outward by the centrifugal barrier and their overlap with the density
rapidly decreases with increasing $\ell$ and in consequence the series 
 (\ref{s2c}) also rapidly converges. The same reasoning repeated for the 
 symmetric wave functions, yields
 \begin{equation}
 C_+(k)=\dfrac{8\pi}{k^2}\sum_{\ell=even} (2\ell+1)\,\int_0^\infty D(r)
 \mid\negthickspace u_\ell(k,r)\negthickspace\mid^2 \D r
 \label{s2e}
\end{equation} 
and for $C_-(k)$  analogous formula holds but the summation
 would be over odd $\ell$.
\par
 As an important example we 
 consider the Coulomb interaction $V_c(r)=\alpha/r$ where $\alpha$
is the fine structure constant times a product of the charge numbers.
The radial  functions are well known
\begin{equation}
  u_\ell(k,r)=F_\ell(\eta,\rho)
 \label{s15}
\end{equation}
with $\rho=kr,\quad \eta=\alpha\mu/k$ where $\mu$ is the reduced mass and the regular
Coulomb functions $F_\ell(\eta,\rho)$ are defined in Abramowitz and Stegun \cite{abram} 
(for numerical methods of computing them cf. \cite{michel} and references therein).
If the two protons experienced only Coulomb interaction, 
the pp correlation function respecting Pauli principle would be 
\begin{equation}
\begin{split}
C_{pp}(k)=  \dfrac{2\pi}{k^2}\left\{
  \sum_{\ell=even}(2\ell+1)\int_0^\infty D(r)\,F_\ell(\eta,kr)^2 \,\D r 
 \right. + \\ \mbox{} + \left.
 3\; \displaystyle
 \sum_{\ell=odd}(2\ell+1)\int_0^\infty D(r)\,F_\ell(\eta,kr)^2 \,\D r
 \right\}
\end{split}
 \label{s15a}
\end{equation} 
where the integrals require numerical treatment.  
\par
 The last step is the inclusion of strong interaction which, in general,  leads to
complications stemming from the fact that the nuclear forces have non-central
components like the tensor interaction or spin-orbit force, and the orbital momentum
is not a good quantum number. 
However, for low energies ($k<100\,MeV/c$, say) the dominant contribution
comes from the s-wave and the interaction in higher partial waves
can be neglected. In this case $\ell$ is still a good quantum number and
including both, the Coulomb and the nuclear s-wave interaction, the pp correlation
function respecting Pauli principle, reads
\begin{equation}
\begin{split}
C_{pp}(k)=  \dfrac{2\pi}{k^2}\left\{
\int_0^\infty D(r) \mid\negthickspace u_0(k,r)\negthickspace\mid^2 \D r+
 \displaystyle \sum_{\ell=even>0} (2\ell+1)\,
\displaystyle \int_0^\infty D(r)\,F_\ell(\eta,kr)^2 \,\D r\right\}+
\\
   \mbox{} +  \dfrac{6\pi}{k^2}       \; \displaystyle 
 \sum_{\ell=odd} (2\ell+1)\,\int_0^\infty D(r)\,F_\ell(\eta,kr)^2 \,\D r
 \hspace*{.04\textwidth}
\end{split}
 \label{d7}
\end{equation}  
where $u_0(k,r)$ is the solution of the wave equation for $\ell=0$ involving both,
the Coulomb and the strong interaction potential. The above formula
constitutes the basis for all our computations.
\par
To get a feeling how (\ref{d7}) works in practice 
it is instructive to provide some example.
Perhaps the simplest simulation of the nuclear force delivers the delta-shell
potential model when $V(r)$ is taken in the form
\begin{equation}
2\mu\, V(r) = -(s/R) \,\delta(r-R)
 \label{d1}
\end{equation}  
where $s$ is a dimensionless strength parameter and $R$ is the range parameter,
assuming that the potential (\ref{d1})  is operative in s-wave only. 
 The wave function $u_0(k,r)$
may be obtained by solving the appropriate Schr\"{o}dinger equation 
but here we shall use for that purpose the equivalent Lippmann-Schwinger equation
which incorporates the correct asymptotic boundary condition. 
More explicitly, we are presented with the integral equation
\begin{equation}
u_0(k,r) = F_0(\eta,kr)+ 2\mu\int_0^\infty g_0^+(r,r^{\,\prime})
\,V(r^{\,\prime})\,u_0(k,r^{\,\prime})\,\D r^{\,\prime}
  \label{d2}
\end{equation} 
where  $g_0^+(r,r^{\,\prime})$ is the outgoing wave Coulomb Green's function
\begin{equation}
g_0^+(r,r^{\,\prime})=-(1/k)\, F_0(\eta,kr_<)\,
 \left[G_0(\eta,kr_>) +\I\, F_0(\eta,kr_>)\right]
 \label{d3}
\end{equation} 
 with $G_\ell(\eta,kr)$
denoting the irregular Coulomb wave function (cf. \cite{abram}, \cite{michel}) and
the other symbols are:  $r_<=\min(r,r');\;r_>=\max(r,r')$. Note, that
 acting with the operator $\D^2/\D r^2 + k^2 - 2\mu\alpha/r$ on both sides
of (\ref{d2}), one immediately recovers the Schr\"{o}dinger equation.
 For the potential (\ref{d1}) it is a trivial matter to obtain the
solution of (\ref{d2}) and the latter is given in an analytic form
 \begin{equation}
u_0(k,r)=
\begin{cases}
 A(k)\,F_0(\eta,kr) \quad & {\rm for}\quad r\leq R \\
 F_0(\eta,kr)+f_0(k)\,k[G_0(\eta,kr) +\I\, F_0(\eta,kr)]      
              \quad & {\rm for} \quad r \geq R
\end{cases}
 \label{d4} 
 \end{equation} 
where the amplitude $ A(k) $ is
\begin{equation}
  A(k) =  \dfrac{1}{1-(s/kR)\,F_0(\eta,kR)\,[G_0(\eta,kR)+\I\,F_0(\eta,kR)]}
  \label{d5}
\end{equation} 
and the second constant
 $f_0(k)$ which is the appropriate pp scattering amplitude, is given as
\begin{equation}
f_0(k) = (s/k^2 R) F_0(\eta,kR)^2\; A(k). 
 \label{d6}
\end{equation} 
Feeding (\ref{d7}) with  $u_0(k,r)$ given in (\ref{d4}), and performing 
(numerically) the integrations we obtain the correlation function.
For ''realistic'' nuclear potentials the procedure would be similar,
 the only difference 
being that $u_0(k,r)$ must be generated numerically from (\ref{d2}).
\begin{figure}[htp]
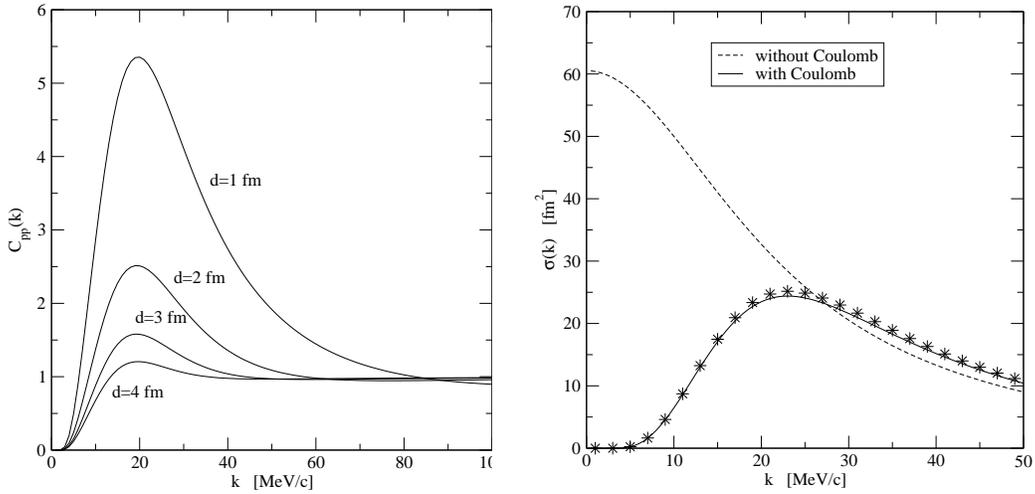

\parbox{0.40\textwidth}{\hspace{-1cm}\includegraphics[width=0.45\textwidth]{plot.eps}}
\parbox{0.40\textwidth}{\includegraphics[width=0.45\textwidth]{sig.eps}}
 \caption{
 (left)
 Two-proton correlation function
 versus $k$ for different source radii calculated from the
 Reid soft core potential.
 (right)
 Partial wave ($\ell=0$) pp cross-section calculated from 
 (\ref{ampl}). The stars show the cross-section computed
 from the toy model formula (\ref{d6}) for $s=0.906$ and $R=1.84$~fm.  
 }  
 \label{fig:0}
\end{figure} 
\par
In our computations we have tried a variety of NN potentials (Reid68,
Reid93, Nijmegen93 and Argonne AV18) but the corresponding
correlation functions were in all cases nearly indistinguishable.
On the other hand, the delta-shell potential resulted in markedly
different shape of the correlation function.     
The plots of the correlation function calculated for 
Reid soft core potential are presented in Fig.\ref{fig:0}(left) for
different radii. In qualitative terms the curves are similar: all show
a depression close to threshold and a prominent peak at about 20~MeV/c.
The same behavior has been observed experimentally \cite{exp}. In our model
the position of the peak does not seem to be dependent upon $d$
whilst its height is very sensitive to the source radius.
The depression close to threshold results from a combined effect of 
Fermi statistics and Coulomb repulsion which both try to keep the two protons
apart causing the correlation to be negative at small $k$. The peak
at 20~MeV/c has a dynamic origin, manifesting strong nuclear attraction
experienced by the protons.  The height of the peak decreases rapidly 
for increasing $d$
 assuming the largest possible value  for $d=0$. 
In the latter case the
source shrinks to a point and the density (\ref{s2})  can be
replaced by a delta function $D(r)=\delta(\bm{r})$. 
The correlation function for a point-like source can be immediately
obtained, and reads
\begin{equation}
\lim_{d\to 0}  C_{pp}(k)= \frac{1}{4}\, \lim_{r \to 0} 
\dfrac{ \mid\negthickspace u_0(k,r)\negthickspace\mid^2}{(kr)^2}=
 \frac{1}{4k^2}\,
  \mid\negthickspace u_0^{\,\prime}(k,0) \negthickspace\mid^2,
 \label{lim}
\end{equation} 
where prime denotes derivative with respect to $r$. 
The expression on the right hand side of (\ref{lim}) is nothing else but
the enhancement factor associated with pp final-state interaction, 
used e.g. to approximate the $pp\to pp\eta$ cross section,
and which is also known to have a prominent peak at 20~MeV/c.
\par
To understand better the origin of the peak at 20 MeV/c let us consider
first a simplified situation with Coulomb interaction switched off.
Close to threshold the NN $\ell=0$ scattering amplitude may be expressed in
terms of the scattering length $a$ as $f_0(k)=(-1/a -\I k)^{-1}$ and
in the complex k-plane this amplitude has a pole at $k_p=\I/a$.
 For $a>0$ the pole is on the physical sheet ($\Im k_p>0$) and we
have a bound state. This is the deuteron case. However, when $a<0$,
the pole is on the non-physical sheet ($\Im k_p<0$) and we are left
with a virtual state.
For the pp pair this would be the particle unstable $^2He$ state.
In either case the NN cross section shows a pronounced peak at threshold
as seen in Fig.\ref{fig:0}(right) (dashed curve).
Putting back the Coulomb interaction and 
including the effective range ($r_0$)
term,  a model independent expression
for the low-energy pp scattering amplitude, takes the form
\begin{equation}
 f_0(k) \approx  \dfrac{C_0^2(\eta)}
 {-1/a+\Half r_0 k^2-2k\eta h(\eta)-\I kC_0^2(\eta)}
\label{ampl}
\end{equation}
where $C_0^2(\eta)=2\pi\eta/[\exp{(2\pi\eta)}-1])$ is the Gamow factor,
 $h(\eta)=\Re\psi(1+\I\eta)-
\log\negthickspace\mid\negthickspace\eta\negthickspace\mid$ 
and $\psi$ denotes the
digamma function \cite{abram}. 
The corresponding partial wave 
cross-section, i.e. $ \mid\negthickspace f_0(k)\negthickspace\mid^2$ 
for $a$=-7.78~fm and $r_0$=2.72~fm is presented in Fig.\ref{fig:0}(right)
by the continuous curve and a similar result would be obtained
from the toy model formula  (\ref{d6}) (depicted by stars in Fig.\ref{fig:0}(right)). 
Owing to the Coulomb corrections, the peak in the
cross section is shifted off-threshold by about 20 Mev/c and its height is depressed
to one third of its original value. This is now the very same peak that 
appears in the correlation function and can be regarded as an
artifact of the  $^2He$ state. It is {\it not} a resonance though 
because the real part of the denominator in (\ref{ampl}) equal
$-1/a+\Half r_0 k^2-2k\eta h(\eta)$ does not vanish,
showing instead only a minimum whose position 
is identical with the position of the peak in Fig.\ref{fig:0}(right).
Actually, given the  low-energy pp scattering parameters,  
the position of the peak could have been predicted by equating to
zero the derivative of the
 real part of the denominator in  (\ref{ampl})  and by 
 solving the resulting equation for $k$.


\end{document}